\documentclass[aps, nofootinbib, preprintnumbers, showpacs,
superscriptaddress, twocolumn]{revtex4}

\usepackage{amssymb}
\usepackage[english]{babel}
\usepackage{bm}
\usepackage{color}
\usepackage{dcolumn}
\usepackage[T1]{fontenc}
\usepackage{graphicx}
\usepackage{ifpdf}
\usepackage[latin9]{inputenc}
\usepackage{mathpazo}
\usepackage{subfigure}
\usepackage{url}

\usepackage{hyperref}

\ifpdf
\else
\usepackage{breakurl}
\fi

\begin{document}

\title{A multi-block infrastructure for three-dimensional
  time-dependent numerical relativity}

\author{Erik Schnetter}
\email{schnetter@cct.lsu.edu}
\affiliation{Center for Computation and Technology, 302 Johnston Hall,
  Louisiana State University, Baton Rouge, LA 70803, USA}
\homepage{http://www.cct.lsu.edu/}
\affiliation{Max-Planck-Institut für Gravitationsphysik,
  Albert-Einstein-Institut, Am Mühlenberg 1, D-14476 Golm, Germany}
\homepage{http://numrel.aei.mpg.de/}

\author{Peter Diener}
\email{diener@cct.lsu.edu}
\affiliation{Department of Physics and Astronomy, 202 Nicholson Hall,
  Louisiana State University, Baton Rouge, LA 70803, USA}
\homepage{http://relativity.phys.lsu.edu/}
\affiliation{Center for Computation and Technology, 302 Johnston Hall,
  Louisiana State University, Baton Rouge, LA 70803, USA}
\homepage{http://www.cct.lsu.edu/}

\author{Ernst Nils Dorband}
\email{dorband@cct.lsu.edu}
\affiliation{Department of Physics and Astronomy, 202 Nicholson Hall,
  Louisiana State University, Baton Rouge, LA 70803, USA}
\homepage{http://relativity.phys.lsu.edu/}
\affiliation{Center for Computation and Technology, 302 Johnston Hall,
  Louisiana State University, Baton Rouge, LA 70803, USA}
\homepage{http://www.cct.lsu.edu/}

\author{Manuel Tiglio}
\email{tiglio@cct.lsu.edu}
\affiliation{Department of Physics and Astronomy, 202 Nicholson Hall,
  Louisiana State University, Baton Rouge, LA 70803, USA}
\homepage{http://relativity.phys.lsu.edu/}
\affiliation{Center for Computation and Technology, 302 Johnston Hall,
  Louisiana State University, Baton Rouge, LA 70803, USA}
\homepage{http://www.cct.lsu.edu/}

\date{2006-06-08}


\preprint{LSU-REL-021606, AEI-2006-008}

\begin{abstract}
  We describe a generic infrastructure for time evolution simulations
  in numerical relativity using multiple grid patches.  After a 
  motivation of this approach, we discuss the relative
  advantages of global and patch-local tensor bases.  We describe both
  our multi-patch infrastructure and our time evolution scheme, and
  comment on adaptive time integrators and parallelisation.  We also
  describe various patch system topologies that provide spherical
  outer and/or multiple inner boundaries.
  
  We employ \emph{penalty} inter-patch boundary conditions, and we
  demonstrate the stability and accuracy of our three-dimensional
  implementation.  We solve both a scalar wave equation on a
  stationary rotating black hole background and the full Einstein
  equations.  For the scalar wave equation, we compare the effects of
  global and patch-local tensor bases, different finite differencing
  operators, and the effect of artificial dissipation onto stability
  and accuracy.  We show that multi-patch systems can directly compete
  with the so-called fixed mesh refinement approach; however, one can
  also combine both.  For the Einstein equations, we show that using
  multiple grid patches with penalty boundary conditions leads to a
  robustly stable system.  We also show long-term stable and accurate
  evolutions of a one-dimensional non-linear gauge wave.  Finally, we
  evolve weak gravitational waves in three dimensions and extract
  accurate waveforms, taking advantage of the spherical shape of our
  grid lines.
\end{abstract}

\maketitle

\section{Introduction}

Many of the spacetimes considered in numerical relativity are
asymptotically flat.  An ideal kind of domain for these has its
boundaries at infinity, and at the same time has to handle
singularities which exist or develop within the domain.  In a
realistic setup, the outer boundary is either placed at infinity,
which is topologically a sphere, or one introduces an artificial outer
boundary at some large distance from the origin.  In both cases,
a sphere is a natural shape for the boundary.

There are several possible ways to deal with singularities.  One of the
most promising is \emph{excision}, which was first used by
J. Thornburg \cite{Thornburg87}, where he acknowledges W. G. Unruh for
the idea.  Excision means introducing a inner boundary, so that the
singularity is not in the computational domain any more.  If done
properly, all characteristic modes  on
this inner boundary are leaving the domain, so that no physical boundary
condition is required.
Seidel and Suen \cite{Seidel92a} applied this idea for the first time
in a spherically
symmetric time evolution.

A well posed initial boundary value problem requires in general smooth
boundaries \cite{Secchi1996a}.
Using spherical boundaries satisfies all conditions
above.  Spherical boundaries have not yet been successfully
implemented in numerical relativity with Cartesian grids.
However, there were many attempts to approximate spherical boundaries.
For example, the Binary Black Hole Grand Challenge Alliance used
\emph{blending} outer boundary conditions \cite{Rezzolla99a,
  Brandt00}, where the blending zone was approximately a spherical
shell on a Cartesian grid.
Excision boundaries often approximate a sphere by having a Lego (or
staircase) shape \cite{Brandt00, Shoemaker2003a, Alcubierre:2004bm,
  Sperhake2005a}.
Some of the
problems encountered with excision are attributed to this staircase
shape.
A spherical boundary would be
smooth in spherical coordinates, but these are undesirable because
they have a coordinate singularity on the $z$ axis.  A multi-block
scheme allows smooth spherical boundaries without introducing
coordinate singularities.

Using multiple grid patches is a very natural thing to do in general
relativity.  When one starts out with a manifold and wants to
introduce a coordinate system, then it is a priori not clear whether a
single coordinate system can cover all the interesting parts of the
manifold.  One usually introduces a set of overlapping maps, each
covering a part of the manifold.  After discretising the manifold, one
arrives naturally at multiple grid patches.

Methods using multiple grid patches in numerical relativity were
pioneered in 1987 by J. Thornburg \cite{Thornburg87, Thornburg93},
where he also introduced excision as inner boundary condition for
black holes.
G\'omez et al.\ \cite{Gomez97} use two overlapping stereographic
patches to discretise the angular direction using the eth formalism.
This was later used by G\'omez et al.\ \cite{Gomez98a} to evolve a
single, non-stationary black hole in a stable manner in three
dimensions with a characteristic formulation.
Thornburg \cite{Thornburg2004:multipatch-BH-excision} evolves a
stationary Kerr black hole in three dimensions using multiple grid
patches using a Cauchy formulation.
Kidder, Pfeiffer, and Scheel \cite{Kidder:2005private} have developed
a multi-patch pseudospectral code to evolve first-order hyperbolic
systems on conforming (neighboring patches share grid points),
touching, and overlapping patches.  Scheel et al.\
\cite{Scheel-etal-2003:scalar-fields-on-Kerr-background} used this
method with multiple radial grid patches to evolve a scalar field on a
Kerr background.  Kidder et al.\ \cite{Kidder:2004rw} used this method
with multiple radial grid patches to evolve a distorted Schwarzschild
black hole.

In this paper, we describe a generic infrastructure for time evolution
simulations in numerical relativity using multiple grid patches, and
we show example applications of this infrastructure.
We begin by defining our notation
and terminology in section \ref{sec:terminology}, where we also
discuss various choices that one has to make when using multi-patch
systems.  We describe our infrastructure in section
\ref{sec:infrastructure} and the patch systems we use in section
\ref{sec:patch-systems}.  We test our methods with a scalar wave
equation on flat and curved backgrounds and with the
full vacuum Einstein
equations in sections \ref{sec:scalar-wave} and \ref{sec:einstein}.
We close with some remarks on future work in section
\ref{sec:conclusions}.

\section{Multi-patch systems}
\label{sec:terminology}

Our main motivation for multi-patch systems is that they provide
smooth boundaries without introducing coordinate singularities.  This
allows us to implement symmetric hyperbolic systems of equations for
well posed initial boundary value problems.  However, multi-patch
systems have three other large advantages when compared to using a
uniform Cartesian grid.

\emph{No unnecessary resolution.}  In multi-patch systems the angular
resolution does not necessarily increase with radius.  An increasing
angular resolution is usually not required, and multi-patch systems
are therefore more efficient by a factor $O(n^2)$ when there are
$O(n^3)$ grid points.

\emph{No CFL deterioration for co-rotating coordinates.}  When a
co-rotating coordinate system is used, the increasing angular
resolution in Cartesian grids forces a reduction of the time step size.
Near the outer boundary, the co-rotation speed can even be
superluminal.  This does not introduce any fundamental problems,
except that the time step size has to be reduced to meet the CFL
criterion.  This makes Cartesian grids less efficient by a factor of
$O(n)$ for co-rotating coordinate systems when there are $O(n^3)$ grid
points.

\emph{Convenient radially adaptive resolution.}  In multi-patch
systems, it is possible to vary the radial resolution without
introducing coordinate distortions
\cite{Thornburg2004:multipatch-BH-excision}.
This is not possible with
Cartesian coordinates.  Fish-eye coordinate transformations
\cite{Baker:2001sf} lead to distorted coordinate systems.
In practise, there
is a limit to how large a fish-eye transformation can be, while there
is no such limit for a radial rescaling in multi-patch systems.

These advantages are so large that we think that fixed mesh
refinement
methods may even be superfluous for many applications in numerical
relativity.  Mesh refinement can be used adaptively, e.g.\ to resolve
shock waves in a star.  It would be difficult to handle this with
adaptive patch systems.  However, mesh refinement is also used
statically to have higher resolution in the centre and lower
resolution near the outer boundary.  This case is very elegantly
handled by multi-patch systems.  Multi-patch systems could also be
used to track moving features, such as orbiting black holes.

\subsection{Terminology}

Methods using multiple grid patches are not yet widely used in
numerical relativity, and this leads to some confusion in terminology.

\begin{figure}
  \subfigure[Touching patches, also known as blocks.]
  {\includegraphics[scale=0.3]{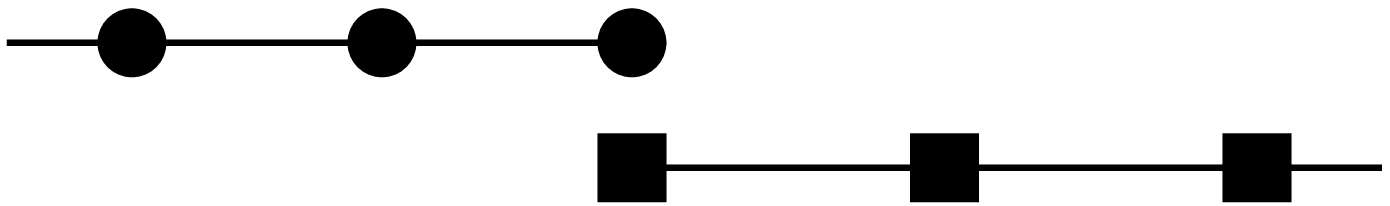}}
  
  \subfigure[Touching patches with additional boundary points.]
  {\includegraphics[scale=0.3]{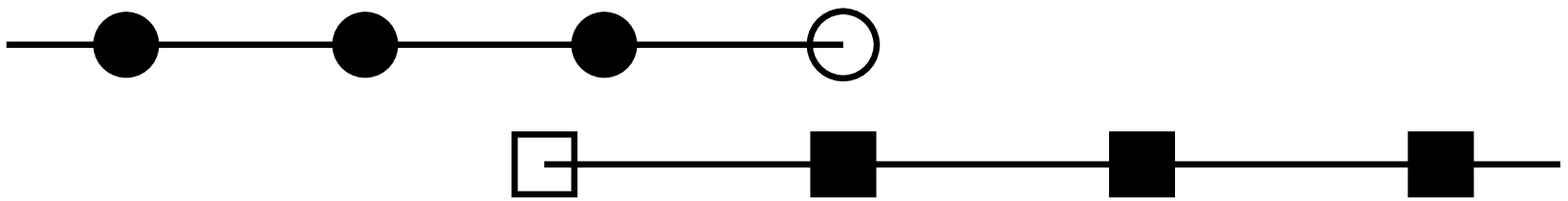}}

  \subfigure[Overlapping patches with boundary points.]
  {\includegraphics[scale=0.3]{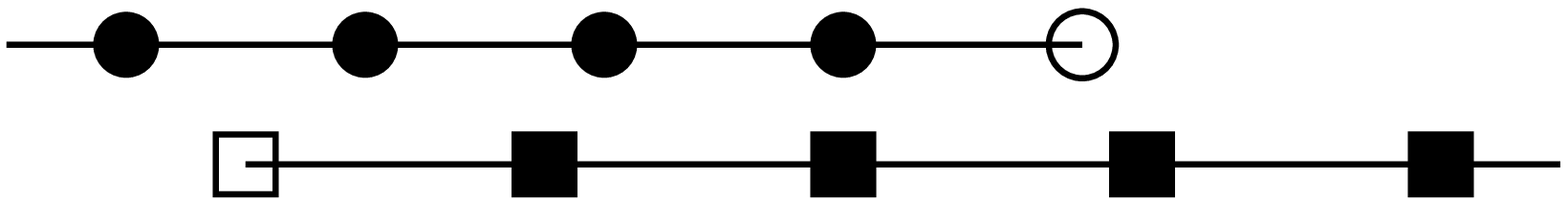}}
  
  \caption{Schematics of the difference between touching and
    overlapping grids for different inter-grid relations.  The black grid
    points are evolved in time, the hollow grid points are determined
    via a boundary condition.}
  \label{fig:grid}
\end{figure}

The notions of multiple patches, multiple blocks, or multiple maps are
all virtually the same.  In differential geometry, one speaks of maps
covering a manifold.  In computational physics, one often speaks of
multi-patch systems when the patches overlap, and of multi-block
systems if they only touch, i.e., if the blocks only have their
boundaries in common.  However, when discretised, there is an
ambiguity as to what part of the domain is covered by a grid with a
certain resolution.  See figure \ref{fig:grid} for an illustration.

In the following, we say that a grid extends from its first to its
last grid point.  This is different from Thornburg's notation
\cite{Thornburg2004:multipatch-BH-excision}: He divides the grid into
\emph{interior} and \emph{ghost} points.%
\footnote{Just to demonstrate that terminology can be really confusing,
  we note that Thornburg's notion of ``ghost points'' is different
  from what Cactus \cite{cactusweb1} calls ``ghost points''.}
The interior points are evolved in time, and a small number of
ghost points are defined through an inter-patch boundary condition,
which is interpolation in his case.  He defines the grid extent as
ranging from the first to the last \emph{interior} point, ignoring the
ghost points for that definition.  Thus, when there are $n$ ghost
points, he calls ``touching'' what we would call an ``overlap of $2n$
points''.  When he speaks of overlapping grids according to his
definition, then there are parts of the domain covered multiple times,
and these overlap regions do not vanish in the continuum limit.

\subsection{Coordinates and tensor bases}

Although not strictly necessary, it is nevertheless very convenient to
have one global coordinate system covering the whole domain.  This
makes it very easy to set up initial conditions from known analytic
solutions, and it also makes it possible to visualise the result of a
simulation.  While a true physics-based visualisation would not rely
on coordinates, currently used methods always display fields with
respect to a coordinate system.

Since our calculations concern quantities that are not all scalars, we
have to make a choice as to how to represent these.  One elegant
solution is to use a tetrad (or triad) formalism, where one represents
vectorial and tensorial quantities by their projections onto the
tetrad elements.  However, this still leaves open the choice of tetrad
elements.

In a multi-patch environment there are two natural choices for a
coordinate basis.  One can either use the global or the patch-local
coordinate system.  Both have advantages and disadvantages.

In a way, using a patch-local coordinate basis is the more natural
choice.  Given that one presumably knows how to evolve a system on a
single patch, it is natural to view a multi-patch system just as a
fancy outer boundary condition for each patch.  In this way, one would
continue to use patch-local coordinates everywhere, while the
inter-patch boundary conditions involve the necessary coordinate
transformations.  It should be noted that these coordinate
transformations mix the evolved variables, because (e.g.\ for a vector
$v_i$) what is $v_1$ on one patch is generally a linear combination of
all $v_i$ in the other patches' coordinate system.

Using a global coordinate basis corresponds to defining a global triad
that is smooth over the entire domain.  This simplifies the
inter-patch boundary conditions significantly, since there is no
coordinate transformation necessary.  Instead, one then has to modify
the time evolution mechanism on the individual patches:

Let the letters $a,b,c\ldots$ denote abstract indices for quantities
in the patch-local tensor basis, and letters $i,j,k\ldots$ abstract
indices in the global tensor basis.  The triad that defines the global
tensor basis is given by $e_a^i$, which is a set of one-forms $e_a$
labelled by a global index $i$.  A vector field $\mathbf{v}$ is
denoted $v^a$ in the patch-local tensor basis and $v^i := e_a^i v^a$
in the global tensor basis.  Note that $v^i$ transforms as a scalar
with respect to the patch-local tensor basis, as a change in the
patch-local tensor basis does not change $v^i$.

When one calculates partial derivatives, e.g.\ through finite
differences, one obtains these always in the patch-local coordinate
basis.  It is then necessary to transform these to the global
coordinate basis.  Since the evolved variables are scalars with
respect to the patch-local coordinate basis, their first derivatives
are co-vectors, and their transformation behaviour is straightforward:
\begin{eqnarray}
  \partial_i & = & \frac{\partial x^a}{\partial x^i}\, \partial_a
\end{eqnarray}
where $\partial x^a/\partial x^i = (\partial x^i/\partial x^a)^{-1} =
(e_a^i)^{-1}$ is the (inverse) \emph{Jacobian} of the coordinate
transformation.  This means that using a global tensor basis
effectively changes the system of evolution equations.

Using a global coordinate basis also has advantages in visualisation.
Visualisation tools generally expect non-scalar quantities to be given
in a global coordinate basis.  Often, one also wants to examine
certain components of non-scalar quantities, such as e.g.\ the radial
component of the shift vector.  When the shift vector is given in
different coordinate bases on each patch, then the visualisation tool
has to perform a non-trivial calculation.

It should be noted that there are many quantities which have a
non-tensorial character, such as e.g.\ the quantities $d_{kij}$ of the
formulation introduced in \cite{Sarbach02b} which we use
below; $d_{kij}$ are partial derivatives of the three-metric.
The quantities $\phi$ and
$\tilde\Gamma^i$ of the BSSN formalism \cite{Alcubierre99d} have an
even more complex transformation behaviour.  $\phi = \ln
\sqrt[12]{\det(\gamma_{ij})}$ is the logarithm of a scalar density,
and $\tilde\Gamma^i = \tilde\gamma^{jk} \tilde\Gamma^i_{jk}$ is a
partial derivative of a tensor density.

It is also necessary to define the set of characteristic variables at
an inter-patch boundary in an invariant manner.  One convenient way to
do so is again to refer to a global coordinate basis.  That is, one
transforms the elements of the state vector to the global coordinate
basis, and can then define the characteristic variables in a unique
way.

Last but not least, there is one more compelling argument for using a
global coordinate basis to represent the state vector.  Since one,
presumably, already knows how to evolve the system within a single
patch, it may be unwise to place all the new complications that a
multi-patch system brings into the inter-patch boundary condition.  By
changing the evolved system to use a global coordinate basis, one
simplifies the inter-patch boundaries significantly, and furthermore
one can implement and test both steps separately.

\section{Infrastructure}
\label{sec:infrastructure}

We base our code on the Cactus framework \cite{Goodale02a, cactusweb1}
using the Carpet infrastructure \cite{Schnetter-etal-03b, carpetweb}.
Cactus is a framework for scientific computing.  As a framework, the
core (``flesh'') of Cactus itself contains no code that does anything
towards solving a physics problem; it contains only code to manage
modules (``thorns'') and let them interact.  Cactus comes with a
set of core thorns for basic tasks in scientific computing, including
time integrators and a parallel driver for uniform grids.  (A driver
is responsible for memory allocation and load distribution on parallel
machines.)

By replacing and adding thorns, we have extended Cactus' capabilities
for multi-patch simulations.  The mesh refinement driver Carpet can
now provide storage, inter-processor communication, and I/O for
multi-patch systems as well, and the multi-patch and mesh refinement
infrastructures can be used at the same time.  The definitions of the
patch systems (see below) and the particular inter-patch boundary
conditions are handled by additional thorns.

\subsection{Infrastructure description}

A computation that involves multiple patches requires implementing
several distinct features.  We decided to split this functionality
across multiple modules, where each module is implemented as a Cactus
thorn.  These are
\begin{description}
\item[Driver (D):] The driver is responsible for memory allocation and
  load distribution, for inter-processor communication, and for I/O.
  It also contains the basic time stepping
  mechanism, ensuring that the application e.g.\ updates the state
  vector on each patch in turn.
\item[Multi-patch system (MP):] The patch system selects how many
  patches there are and how they are connected, i.e., which face is
  connected to what face of which other patch, and whether there is a
  rotation or reflection necessary to make the faces match.  The patch
  system also knows where the patches are located in the global
  coordinate space, so that it can map between patch-local and global
  coordinates.
\item[Penalty boundaries (P):] This thorn applies a penalty boundary
  condition to the right hand side (RHS) of the state vector of one
  face of one patch.  We have described the details in \cite{Diener05b}.
  It calls other routines to convert the state
  vector and its RHS on the patch and on its neighbour to and from
  their characteristic variables; it is thus independent of the
  particular evolution system.
\item[Finite differences (FD):] As a helper module, we implemented
  routines to calculate high order finite differences on the patches
  \cite{Diener05b}.  These operators use one-sided differencing
  near the patch boundaries.
\end{description}

Additionally, we make use of the following features that Cactus
provides:
\begin{description}
\item[Time integrator (TI):] The time integrator calls user-provided
  routines that evaluate the RHS of the state vector, advances the
  state vector in time, and calls boundary condition routines.
\item[Boundary conditions (BC):] The boundary condition infrastructure
  keeps track of what boundary conditions should be applied to what
  faces and to what variables.  It distinguishes between
  \emph{inter-processor boundaries} (which are synchronised by the
  driver), \emph{physical boundary conditions} (where the user applies
  a condition of his/her choice), and \emph{symmetry boundary
    conditions} (which are determined through a symmetry of the
  computational domain, e.g.\ a reflection symmetry about the
  equatorial plane).  We have extended the notion of symmetry
  boundaries to also include inter-patch boundaries, which are in our
  case handled by the penalty method.
\end{description}

Together, this allows multi-patch systems to be evolved in Cactus
within the existing infrastructure.  Existing codes, which presumably
only calculate the right hand side (RHS)
and apply boundary conditions, can make use of
this infrastructure after minimal changes, and after e.g.\ adding
routines to convert the state vector from and to the characteristic
modes.  Existing codes which are not properly modular will need to be
restructured before they can make use of this multi-patch
infrastructure.

We use the penalty method to enforce the inter-patch boundary
conditions.  The penalty method for finite differences is described in
\cite{Carpenter1994a, Carpenter1999a, Nordstrom2001a}, and we describe
our approach and notation in \cite{Diener05b}.

In order to treat systems containing second (or higher) temporal
derivatives with our current infrastructure, one needs to rewrite them
to a form where only contain only first temporal derivatives by
introducing auxiliary variables.  This is always possible.  Strongly
or symmetric hyperbolic systems containing second (or higher) spatial
derivatives \cite{Kreiss:2001cu, Sarbach02a, Nagy:2004td,
  Beyer:2004sv, Gundlach:2005ta, Babiuc:2006wk} could in principle
also be handled without reducing them to first order.  The definition
of the characteristic modes has then to be adapted to such a
formulation, and may then contain derivatives of the evolved
variables.

This multi-patch approach is not limited in any way to finite
differencing discretisations.  It would equally be possible to use
e.g.\ spectral methods to discretise the individual patches (as was
done in
\cite{Scheel-etal-2003:scalar-fields-on-Kerr-background}.)  It may
even make sense to use structured finite element or finite volume
discretisations on the individual patches, using a multi-patch system
to describe the overall topology.

\subsection{Initial condition, boundary conditions, and time
  evolution}

We now describe how the initial conditions are set up and how the state
vector is evolved in time.  This explains in some more detail how the
different modules interact, and what parts of the system have to be
provided by a user of this multi-patch infrastructure.  Each step is
marked in parentheses (e.g.\ ``(MP)'') with the module that performs
that step.

Setting up the initial conditions proceeds as follows:
\begin{enumerate}
\item \emph{Initialise the patch system.}  (MP) The patch system is
  selected, and the location and orientation of the patches is
  determined depending on run-time parameters.  If desired, the patch
  system specification is read from a file (see below).
\item \emph{Set up the global coordinates.}  (MP) For all grid points
  on all patches, the global coordinates are calculated.  At this time
  we also calculate and store the Jacobian of the coordinate
  transformation between the global and the patch-local coordinate
  systems.  If necessary, we also calculate derivatives of the
  Jacobian.  Derivatives may be required to transform non-tensorial
  quantities when a patch-local tensor
  basis is used.  The Jacobian can be calculated either
  analytically (if the coordinate transformation is known
  analytically) or numerically via finite differences.
\item \emph{Initialise the three-metric.}  (User code) Even when the
  evolved system does not contain the Einstein equations, we decided
  to use a three-metric.  This is a convenient way to describe the
  coordinate system, which ---even if the spacetime is flat--- is
  non-trivial in distorted coordinates.  For example, polar-spherical
  coordinates can be expressed using the three-metric $\gamma_{ij} =
  \mathrm{diag} \left( 1, r^2, r^2\sin^2\theta \right)$, and doing so
  automatically takes care of all geometry terms.

  This step is performed by the user code, and it is not necessary to
  use an explicit three-metric in order to use this infrastructure.
  At this time, we initialise the metric at the grid point locations,
  specifying its Cartesian components in the \emph{global} tensor
  basis.
\item \emph{Initialise the state vector.}  (User code) Here we
  initialise the state vector of the evolution system, also again
  specifying its Cartesian components in the \emph{global} tensor
  basis.  When evolving Einstein's equations, this step and the
  previous are combined.
\item \emph{Convert to local tensor basis (if applicable)} (User code)
  It is the choice of the user code whether the state vector should be
  evolved in the global or in the patch-local tensor bases.  We have
  discussed the advantages of either approach above.  If the evolution
  is to be performed in patch-local coordinates, then we transform the
  three-metric and the state vector at this time.  Note that this
  transformation does not require interpolation, since we evaluated
  the initial condition already on the grid points of the individual
  patches.
\end{enumerate}
At this stage, all necessary variables have been set up, and the state
vector is in the correct tensor basis.

The time evolution steps occur in the following hierarchical manner:
\begin{enumerate}
\item \emph{Loop over time steps.}  (D) The driver performs time steps
  until a termination criterion is met.  At each time step, the time
  integrator is called.
  \begin{enumerate}
  \item \emph{Loop over substeps.}  (TI) We use explicit time
    integrators, which evaluate the RHS multiple times and calculate
    from these the updated state vector.
    \begin{enumerate}
    \item \emph{Evaluate RHS.}  (User code) The RHS of the state
      vector is calculated, using e.g.\ the finite differencing thorn.
    \item \emph{Apply boundary conditions to RHS.}  (BC) We decided to
      apply boundary conditions not to the state vector, but instead
      to its RHS.  This is necessary for penalty boundaries, but is a
      valid choice for all other boundaries as well.  In our
      simulations, we do not apply boundary conditions to the state
      vector itself, although this would be possible.  Both the
      inter-patch and the outer boundary conditions are applied via
      the multi-patch infrastructure in a way we explain below.
    \item \emph{Update state vector.}  (TI) Calculate the next ---or
      the final--- approximation of the state vector for this time step.
    \item \emph{Apply boundary conditions to the state vector.}  (User
      code) In our case, nothing happens here, since we apply the
      boundary conditions to the RHS of the state vector instead.
    \end{enumerate}
  \item \emph{Analyse simulation state.}  (D) The driver calls various
    analysis routines, which e.g.\ evaluate the constraints, or
    calculate the total energy of the system, or output quantities to
    files.
  \end{enumerate}
\end{enumerate}

The multi-patch thorn knows which faces of which patches are
inter-patch boundaries and which are outer boundaries.
Inter-processor boundaries are handled by the driver and need not be
considered here.  Symmetry boundary conditions (such as e.g.\ a
reflection symmetry) are currently not implemented explicitly, but
they can be trivially simulated by connecting a patch face to itself.
This applies the symmetry boundary via penalties, which is numerically
stable, but differs on the level of the discretisation error from an
explicitly symmetry condition.

The boundary conditions are applied in the following way:
\begin{enumerate}
\item \emph{Loop over all faces of all patches.}  (MP) Traverse all
  faces, determining whether this face is connected to another patch
  or not.  Apply the boundary condition for this face, which is in our
  case always a penalty condition.
  \begin{enumerate}
  \item \emph{Apply a penalty to the RHS.}  (P) Apply a penalty term
    to the RHS of all state vector elements.  This requires
    calculating the characteristic variables at the patch faces.
    \begin{enumerate}
    \item \emph{Determine characteristic variables.}  (User code)
      Calculate the characteristic variables from the state vector on
      the patch to which the penalty should be applied.  If applying
      an inter-patch boundary condition, calculate the characteristic
      variables from the other patches' state vector as well.  If this
      is an outer boundary, then specify characteristic boundary data.
    \item \emph{Apply penalty.}  (P) Apply the penalty correction to
      the characteristic variables.
    \item \emph{Convert back from characteristic variables.}  (User
      code) Convert the characteristic variables back to the RHS of
      the state vector.
    \end{enumerate}
  \end{enumerate}
\end{enumerate}

The edges and corners of the patches require some care.  In our
scheme, the edges and corners of the patches have their inter-patch
condition applied multiple times, once for each adjoining patch.
In the case of penalty boundary conditions, the edge and corner grid
points are penalized multiple times, and these penalties are added up.
This happens for each patch which shares the corresponding grid
points.  We have described this in more detail in \cite{Diener05b}.

For nonlinear equations, there is an ambiguity in the definitions of
the characteristic variables and characteristic speeds on the
inter-patch boundaries.  Since we use the penalty method, the state
vector may be different at the boundary points on both sides of the
interface.  When the metric is evolved in time, then it is part of the
state vector, and it may be discontinuous across the interface.  The
definition of the characteristics depends on the state vector in the
nonlinear case.  It can thus happen that the characteristic speeds are
all positive when calculated at the boundary point on one side of the
interface, and all negative when calculated on the other side of the
interface.  One has to pay attention to apply the penalty terms in a
consistent manner even if this is the case.  In our scheme (as
described above), we always calculate the characteristic information
using the state vector on that side of the interface to which the
penalty terms are applied.  This scheme does not prefer either side of
the interface, but it is not fully consistent for nonlinear equations.

Instead of using penalty terms to apply boundary conditions, one could
also apply boundary conditions in other ways, e.g.\ through
interpolation from other patches, or by specifying Dirichlet or von
Neumann conditions.  Our infrastructure is ready to do so, but we have
not performed a systematic study of the relative advantages of e.g.\
penalty terms vs.\ interpolation.

\subsection{Time integration}

It is common in numerical relativity to use explicit time integration
methods.  These limit the time step size to a certain multiple of the
smallest grid spacing in the simulation domain.  In non-uniform
coordinate systems, one has to determine the smallest grid spacing
explicitly, since it is the proper distance between neighbouring grid
points that matters, not the grid spacing in the patch-local
coordinate systems, and the proper distance can vary widely across a
patch.  If the three-metric, lapse, or shift are time-dependent, then
the proper distances between the grid points also vary with time.

Furthermore, the maximum ratio between the allowed time
step size and the grid spacing depends not only on the system of
evolution equations; it depends also on the spatial discretisation
operators that are used, on the amount of artificial dissipation, and
on the strength of the penalty terms.  While all this can in principle
be calculated a priori, it is time consuming to do so.

Instead, we often use adaptive time stepping,
for example using the adaptive step
size control of the Numerical Recipes \cite[chapter 16.2]{Press86}.
One can specify a time integration accuracy that is much higher than
the spatial accuracy, and thus obtain the convergence order according
to the spatial discretisation.  In practise, one would rather specify
a time integration accuracy that is comparable to the spatial
accuracy.  Adaptive time stepping would also have the above advantages
when used on a single Cartesian grid.

We would like to remark on a certain peculiarity of adaptive time
stepping.  With a fixed time step size, instabilities manifest
themselves often in such a way that certain quantities grow without
bound in a finite amount of simulation time.  Numerically, one notices
that these quantities become infinity or nan (not a number) at some
point when IEEE semantics \cite{Goldberg91} are used for floating
point operations.  With an adaptive step size, this often does not
happen.  Instead, the step size shrinks to smaller and smaller values,
until either the step size is zero up to floating point round-off error, or
the time integrator artificially enforces a certain, very small
minimum step size.  In both cases, computing time is used without
making any progress.  This case needs to be monitored and detected.

\subsection{Parallelisation}

Our current implementation parallelises a domain by splitting and
distributing each patch separately onto all available processors.
This is not optimal, and it would be more efficient to split patches
only if there are more processors than patches, or if the patches have
very different sizes.  This is a planned future optimisation.

\begin{figure}
  \includegraphics[width=0.45\textwidth]{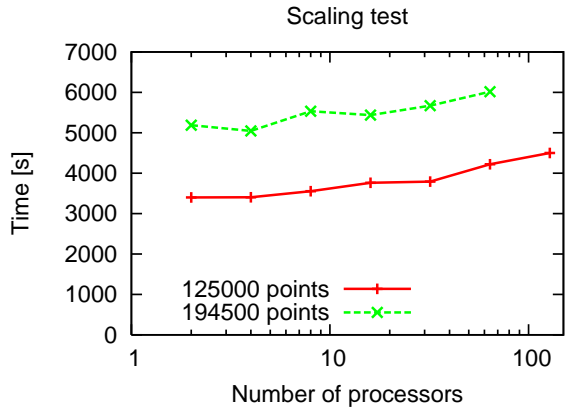}
  \caption{Wall clock time vs.\ numbers of processors for $100$ time
    steps of a test problem.  We keep the load of each processor
    approximately constant at $125000$ and $194500$ grid points,
    respectively.  Our implementation scales up to at least 128
    processors.}
  \label{fig:scaling}
\end{figure}

We have performed a scaling test on multiple processors.  We solve a
simple test problem on a patch system with multiple patches and
measure the time it takes to take 100 time steps.%
\footnote{This test was performed
  with a 4th order Runge--Kutta integrator, the
  scalar wave equation formulated in a patch-local tensor basis, a
  seven-patch system, the $D_{6-5}$ differencing operators, and a
  Mattsson--Sv\"ard--Nordstr\"om dissipation operator.  We varied the
  number of grid points per patch from $65^3$ to $253^3$ to keep the
  load per processor approximately constant.  See section
  \ref{sec:scalar-wave-minkowski} below, where these details are
  explained.}
As we increase the number of processors, we also increase the number
of grid points, so that the load per processors remains approximately
constant.  This is realistic, because one chooses the number of
processors that one uses for a job typically depending on the problem
size.  Figure \ref{fig:scaling} shows the results of the scaling tests
for two such problem sizes.  We find that our implementation scales
well up to at least 128 processors, and would probably continue to
scale to larger numbers.  See \cite{cactusbenchmarkweb} for a
comparison of other benchmarks using Cactus and Carpet.

It would also be possible to distribute the domain onto the available
processors by giving (at least) one domain to each processor.  This
would mean that one splits domains when one adds more processors,
introducing additional inter-patch boundary conditions.  Penalty
inter-patch boundary conditions are potentially more efficient than
using ghost zones, since they require an overlap of only one single
grid point.  An $n$th order accurate finite differencing scheme, on
the other hand, requires in general an overlap of $2n$ grid points.
Penalty boundary conditions thus require less communication between
the patches.  A disadvantage of this scheme is that the exact result
of a calculation then depends on the number of processors.  Of course,
these differences are only of the order of the discretisation error.

Such differences are commonly accepted when e.g.\ elliptic equations
are solved.  Many efficient algorithms for solving elliptic equations
apply a domain decomposition, assigning one domain to each processor,
and using different methods for solving within a domain and for
coupling the individual domains.  The discretisation error in the
solution depends on the number of domains.  For hyperbolic equations
that are solved with explicit time integrators, it is often customary
to not have such differences.  On one hand, this may not be necessary
to achieve an efficient implementation, and on the other hand, it
simplifies verifying the correctness of a parallel implementation if
the result is independent of the number of processors.  However, there
are no fundamental problems in allowing different discretisation
errors when solved on different numbers of processors, especially if
this may lead to a more efficient implementation.

\section{Patch systems}
\label{sec:patch-systems}

We have implemented a variety of patch systems, both for testing and
for standard application domains.  It is also possible to read patch
systems from file.

Simple testing patch systems are important not only while developing
the infrastructure itself, but also while developing applications
later on.  Since the application has to provide certain building
blocks, such as e.g.\ routines that convert to and from the
characteristic representation, it is very convenient to test these in
simple situations.  Many patches have distorted local coordinate
systems, and it is therefore also convenient to have patches with
simple (one-dimensional) coordinate distortions.  This can be used to
test the tensor basis transformations --- keeping in mind that some
variables will not be tensorial, but will rather be tensor densities,
or partial derivatives of tensors, with correspondingly more involved
transformation behaviours.

\begin{figure}
  \includegraphics[width=0.45\textwidth]{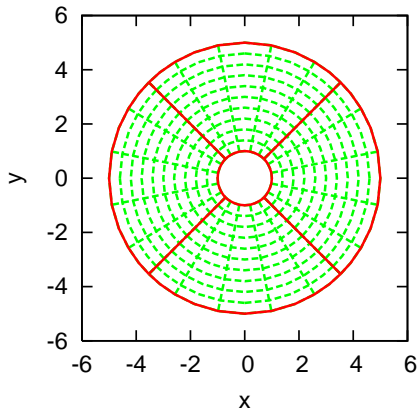}
  \caption{\label{fig:six-patches}%
    A cut in the equatorial plane of six patches, in which four
    patches are visible.  The outer and inner domain boundaries are
    spheres.  There is one radial coordinate spanning $r=\text{const}$
    surfaces, and two angular coordinates perpendicular to that.  The
    radial coordinate is smooth across patch boundaries.}
\end{figure}

We have currently two types of realistic patch systems implemented:

\paragraph{Six patches:}
This system consists of six patches that cover a spherical shell,
i.e., a region with $r_{\text{min}} \le r \le r_{\text{max}}$.  We use
the same patch-local coordinates as in \cite{Lehner2005a} and
\cite{Diener05b}.
This system is useful if the origin is not part of the domain, e.g.\
for a single black hole.  See figure \ref{fig:six-patches} for an
illustration.

\begin{figure}
  \includegraphics[width=0.45\textwidth]{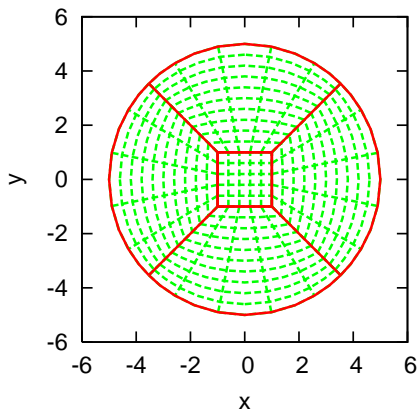}
  \caption{\label{fig:seven-patches}%
    A cut in the equatorial plane of seven patches, in which five
    patches are visible.  The outer boundary is a sphere, the inner
    patch is a cube.  There is again one radial coordinate, but it
    does not span $r=\text{const}$ surfaces and it is not smooth
    across patch boundaries except at the outer boundary.  The two
    angular coordinates are the same as in the six-patch system.}
\end{figure}

\paragraph{Seven patches:}
This system consists of one cubic patch that covers the region near
the origin, and six additional patches that cover the exterior of the
cube until a certain radius $r_{\text{max}}$.  We use the same
patch-local coordinates as in \cite{Diener05b}, which are derived
from the six-patch coordinates above.
This system is useful if the origin
should be part of the domain, e.g.\ for a single neutron star.  See
figure \ref{fig:seven-patches} for an illustration.

In addition to these two types, we have variations thereof, e.g.\ a
system consisting of only one of the six patches assuming a sixfold
reflection symmetry.  We have individual patch types as generic
building blocks, and we can glue them together to form arbitrary patch
systems.

After seeking input from the computational fluid dynamics community,
where multi-block systems are commonly used to obtain body-fitted
coordinate systems, we decided that setting up patch systems by hand
is too tedious, and that commercial tools should be used for that
instead.%
\footnote{We are indebted to our esteemed colleague F. Muldoon for
  teaching us about the state of the art in grid generation.}
We therefore implemented a patch system reader that understands the
GridPro \cite{gridproweb} data format.  This is a straightforward
ASCII based format which is specified in the GridPro documentation,
and support for other data formats could easily be
implemented as well.

\begin{figure}
  \includegraphics[width=0.45\textwidth]{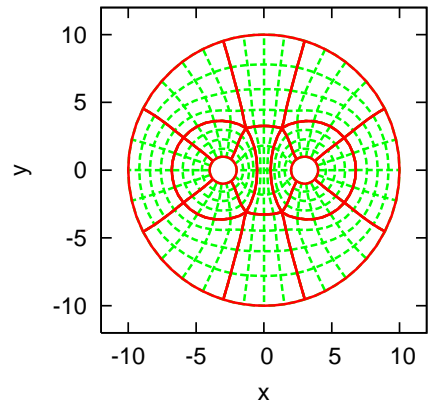}
  \caption{\label{fig:two-black-holes}%
    A cut in the equatorial plane of the imported binary black hole
    patch system.  The outer and inner boundaries are spheres.  Near
    the boundaries, the coordinate system is similar to spherical
    coordinates, i.e., there is one coordinate direction
    perpendicular to and two direction tangential to the boundary.}
\end{figure}

\begin{figure}
  \subfigure[whole domain]
  {\includegraphics[width=0.45\textwidth]{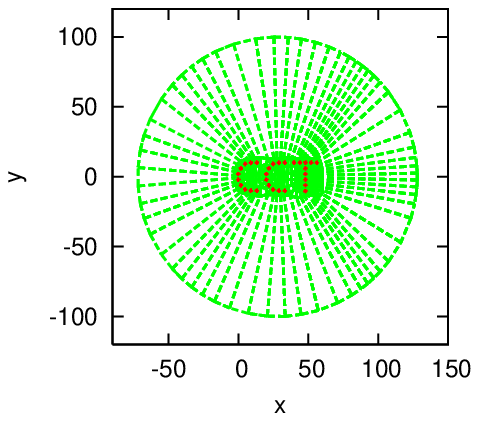}}

  \vspace{-1cm}
  \subfigure[region near the origin]
  {\includegraphics[width=0.45\textwidth]{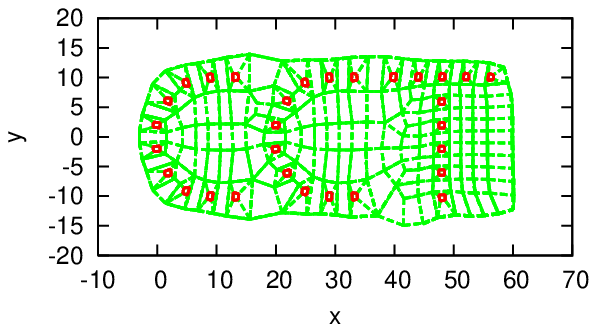}}
  \caption{A cut in the equatorial plane of the imported demonstration
    patch system with 30 spherical holes in arbitrary positions close
    to the centre.  The resolution in the centre is much higher than
    near the outer boundary, demonstrating how to achieve the same
    effect as fixed mesh refinement with multiple patches.}
  \label{fig:cct}
\end{figure}

Using GridPro, we could easily import patch systems with two holes and
27 patches (for a generic binary black hole system; see figure
\ref{fig:two-black-holes}) and with e.g.\ 30 holes and 865 patches
(for demonstration purposes; see figure \ref{fig:cct}).
Another advantage of a tool like GridPro is that the grid points are
automatically evenly distributed over the domain, which may be
difficult to ensure if the grid is constructed by hand.
%
%%\begin{figure}
%%  \includegraphics[width=0.45\textwidth]{runs/cct}
%%\end{figure}
%

\section{Tests with the scalar wave equation}
\label{sec:scalar-wave}

We test our multi-patch infrastructure with a scalar wave equation
on an arbitrary, time-independent background.  Since we express the
coordinate distortions via a generic three-metric, there is ---from
the point of view of the code--- no difference between a flat and a
curved spacetime.  A stationary black hole background requires
non-trivial lapse and shift functions, but these are desirable even
for flat spacetimes: a non-zero shift makes for moving or rotating
coordinate systems (implementing fictitious forces), and a non-unity
lapse could be used to advance different parts of the domain with
different speeds in time, which can improve time integration
efficiency (although we did not use it for that purpose).

We use the notation $\alpha$ for the lapse, $\beta^i$ for the shift
vector, $\gamma_{ij}$ for the three-metric, $\gamma^{ij}$ for its
inverse, and $\gamma = \det (\gamma_{ij})$ for its determinant.

We evolve the scalar wave equation
\begin{eqnarray}
  \label{eqn:sw}
  \Box u & = & 0
\end{eqnarray}
by introducing the auxiliary variables
\begin{eqnarray}
  \label{eqn:rho}
  \rho & = & \partial_t u \\
  \label{eqn:v}
  v_i & = & D_i u \text{.}
\end{eqnarray}
This renders the system into a first order form, leading to the time
evolution equations
\begin{eqnarray}
  \label{udot}
  \partial_t u & = & \rho \\
  \label{rhodot}
  \partial_t \rho & = & \beta^i \partial_i \rho +
  \frac{\alpha}{\sqrt{\gamma}} \partial_i
  \left[ \frac{\sqrt{\gamma}}{\alpha} \left( \beta^i \rho + \alpha^2
      H^{ij} v_j \right) \right] \\
  \label{vdot}
  \partial_t v_i & = & \partial_i \rho
\end{eqnarray}
with
\begin{eqnarray}
  H^{ij} & = & \gamma^{ij} - \beta^i \beta^j / \alpha^2 \text{.}
\end{eqnarray}
If discretised with operators that satisfy summation by parts, this
system is numerically stable with respect to the energy
\begin{eqnarray}
  E & = & \frac{1}{2} \int \frac{\sqrt{\gamma}}{\alpha} \left[ \rho^2
    + \alpha^2 H^{ij} v_i v_j \right]\, dV
\end{eqnarray}
for $|\beta|<\alpha$.  In this case, this system is symmetric
hyperbolic, and its characteristic variables and speeds are listed in
\cite{Lehner2005a}.

We present below evolutions of the scalar wave equation on a flat
background with the seven patch system and on a Kerr--Schild
background with the six patch system.  We compare the respective
benefits of using a global or a patch-local tensor basis, and we study
the behaviour of scalar waves on a Kerr--Schild background as a test
problem.

\subsection{Comparing global and patch-local tensor bases}
\label{sec:scalar-wave-minkowski}

We present here time evolutions of the scalar wave equation on a flat
background with the seven-patch system.  We compare two formulations,
one based on a global, the other based on a patch-local tensor basis.
We also apply a certain amount of artificial dissipation to the
system, which is necessary because our formulation has non-constant
coefficients.  We use two different kinds of artificial dissipation,
which were introduce by Kreiss and Oliger \cite{Kreiss73} and by 
Mattsson, Sv\"ard, and Nordstr\"om \cite{Mattsson2004a}, respectively.

We set the initial condition from an analytic solution of the wave
equation, namely a traveling plane wave.  We also impose the analytic
solution as penalty boundary condition on the outer boundaries.  This
is in the continuum limit equivalent to imposing no boundary condition
onto the outgoing characteristics and imposing the analytic solution
as Dirichlet condition onto the incoming characteristics.  The patch
system has the outer boundary at $r=3$, and the inner, cubic patch has
the extent $[-1;+1]$.  We use the penalty strength $\delta=0$
% Erik checked this number
at the inter-patch and at the outer boundaries.
See \cite{Diener05b} for our notation for the penalty terms.

The traveling plane wave is described by
\begin{eqnarray}
  u(t,x_i) & = & A\, \cos \left[ 2\pi \left( k_i x^i + \omega t
    \right) \right]
\end{eqnarray}
with $\omega^2 = \delta^{ij} k_i k_j$.  We set $A=1$ and
$k_i=[0.2,0.2,0.2]$, so that the wave length is $5/\!\sqrt{3}$.  We
construct the solutions for $\rho$ and $v_i$ from $u$ via their
definitions (\ref{eqn:rho}) and (\ref{eqn:v}), and evaluate these at
$t=0$ to obtain the initial condition.  The flat background has
$\alpha=1$, $\beta^i=0$, and $\gamma_{ij}=\delta_{ij}$ in the global
tensor basis; the patch-local metric is constructed from that via a
coordinate transformation.

We use dissipation operators that are compatible with summation by
parts (SBP) finite difference operators.  Introduced by
Mattsson, Sv\"ard, and Nordstr\"om in \cite{Mattsson2004a}, we call them
``MSN'' operators.  They are constructed according to
\begin{eqnarray}
  A_{2p}^\mathrm{MSN} & = & - \frac{\epsilon}{2^{2p}}\, h^{2p}\;
  \Sigma^{-1} D_p^T B_p D_p \text{,}
  \label{eqn:msn-disspation}
\end{eqnarray}
where $2p$ is the order of the interior derivative operator,
$\epsilon$ is the dissipation strength, $h$ is the grid spacing,
$\Sigma$ is the norm with respect to which the derivative operator
satisfies SBP, $D_p$ is a consistent approximation of a $p$th
derivative, and $B_p$ is a diagonal matrix.  The scaling with grid spacing
is in contrast to standard Kreiss--Oliger (KO) dissipation operators
\cite{Kreiss73}, where
the scaling $h^{2p-1}$ is used.  Experience has shown that with the MSN
scaling it is sometimes necessary to increase the dissipation strength
when resolution is increased in order to maintain stability.  For this
reason we have implemented SBP compatible KO dissipation
operators constructed according to
\begin{eqnarray}
  A_{2p}^\mathrm{KO} & = & - \frac{\epsilon}{2^{2p+2}}\, h^{2p+1}\; \Sigma^{-1}
  D_{p+1}^T B_p D_{p+1} \text{,}
  \label{eqn:ko-disspation}
\end{eqnarray}
where as before $\Sigma$ is the norm of the $2p$th order accurate SBP
derivative operator.  This yields a dissipation operator with
KO scaling that has the same accuracy as the SBP
derivative operator near the boundary (and one order higher in the
interior).  The price is having to use a slightly wider stencil.

We use $21$ grid points in the angular and in the radial directions.
The central patch also has $21$ grid points in each direction.
This is a very coarse resolution.  We use the $D_{6-5}$ stencil of
\cite{Diener05b}, which is globally sixth order accurate, and add
compatible artificial dissipation to the system of both MSN and KO type
as described above.  We choose a dissipation coefficient $\epsilon=3.0$.
We use a transition region that is $0.3$ times the size of the patch.  The
overall system is then sixth order accurate.

\begin{figure}
  \includegraphics[width=0.45\textwidth]{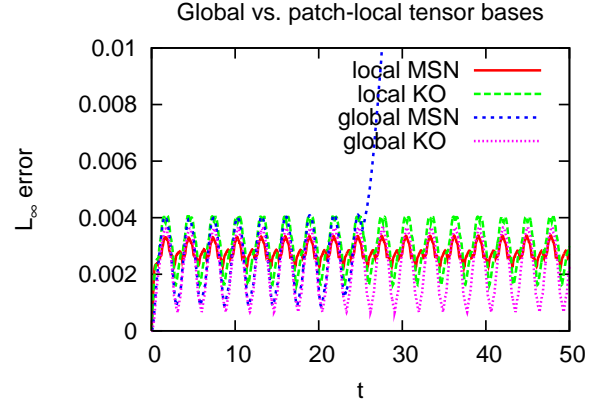}
  % this has 20 cells, 6-5
  \caption{Comparing global and patch-local tensor bases, and
    Mattsson--Sv\"ard--Nordstr\"om (MSN) and Kreiss--Oliger (KO)
    dissipation operators.  The graphs show the $L_\infty$ norm of the
    solution error vs.\ time for a coarse resolution on a seven-patch
    system.  Some artificial dissipation is necessary to stabilise the
    system, since it has non-constant coefficients.  For this
    particular value of the dissipation strength $\epsilon$, using a
    global tensor basis is unstable with the MSN dissipation
    operators, but stable with the KO operators.  With higher values
    of $\epsilon$, the system is stable for both dissipation
    operators.}
  \label{fig:localglobal}
\end{figure}

With a patch-local tensor basis and diagonal norm operators the system
is strictly stable, i.e., the numerical error is at any given resolution
bounded (up to boundary terms) by a constant (see also \cite{Diener05b}),
while with a global tensor basis, a small amount of artificial dissipation
is required.  However, since we are using restricted full norm operators for
this test, dissipation is required for both the patch-local and patch-global
case.

Figure \ref{fig:localglobal} shows the $L_\infty$ norm of the solution
error vs.\ time up to $t=50$.  The discretisation using
MSN dissipation is unstable for $\epsilon=3.0$ when a global tensor basis is
used, but it is stable when a local tensor basis is used.  Larger values of
$\epsilon$ also stabilise the global tensor basis discretisation.  For the
KO dissipation, a dissipation strength $\epsilon = 3.0$ is
sufficient to stabilise both the local and global tensor basis
formulations.
Note that the error levels are very similar in all cases, showing that the main 
difference between the patch-local and patch-global tensor basis 
implementations is that more dissipation is necessary in order to stabilise
the system.

\begin{figure}
  \includegraphics[width=0.45\textwidth]{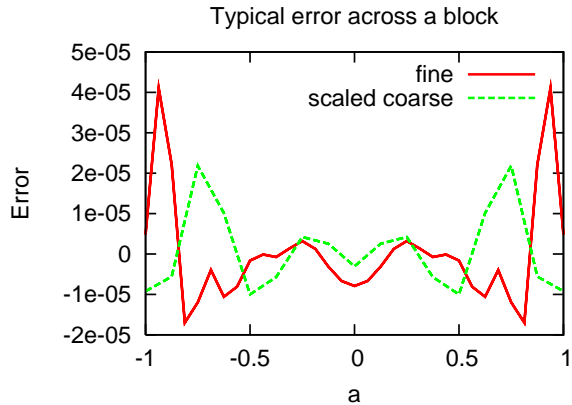}
  % the errors are for a local basis, KO diss 0.4, 32 cells, 6-5
  \caption{This figure shows typical solution error shapes across a
    block.  It shows the error along the $a$ coordinate line of the
    $+x$ patch of a seven-patch system.  This coordinate line
    corresponds to a $\phi$ coordinate line in spherical coordinates,
    but has a kink at the patch interface; see figure
    \ref{fig:seven-patches}.  This figure shows two resolutions which
    differ by a factor of two, and the coarse error is scaled
    according to 6th order convergence.  The error is largest near the
    block boundary.  Because the SBP stencils of the $D_{6-5}$
    operator are modified on 7 grid points near each boundary,
    convergence near the boundary is not obvious from this graph.  We
    show 6th order convergence for this case in \cite{Diener05b}.}
  \label{fig:patcherror}
\end{figure}

Finally, we show a typical shape for solution errors in figure
\ref{fig:patcherror}.  This shows the solution errors in the quantity
$u$ across the $+x$ block for two different resolutions.  The
simulation started with a Gaussian pulse as initial condition.  The
graph shows the errors at the time $t=25$ along the $a$ coordinate
line for $b=0,\,c=0$; this coordinate line is approximately the $\phi$
coordinate line, except that it has a kink at the block interfaces.
(See figure \ref{fig:seven-patches} for an illustration.)  $b=0$
places the coordinate line into the equatorial plane, and $c=0$
chooses the centre of the block in the radial direction.  The coarse
block had $17\times 17\times 141$ grid points with the outer boundary
at $R=15$, and the fine block had twice this resolution.  The
simulation was run with the $D_{6-5}$ operator, and we expect 6th
order convergence.  We have intentionally chosen rather coarse
resolutions in the angular direction.  Because the SBP stencils of the
$D_{6-5}$ operator are modified on 7 grid points near each boundary,
convergence at the boundary is not obvious from this graph.  However,
we show in \cite{Diener05b} that this system converges indeed to
6th order.

\subsection{Scalar wave equation on a Kerr--Schild background}

We also present time evolutions of the scalar wave equation on a
Kerr--Schild \cite[section 3.3]{Cook00a1} background with six patches,
which we use to excise the singularity.  We choose the mass $M=1$ and
the spin $a=0.9$ for the background.  We place the inner boundary at
$r=1.4$ and the outer boundary at $r=201.4$.

We use as initial condition $u(0,x^i)=0$, $v_i(0,x^i)=0$, and a
modulated Gaussian pulse
\begin{eqnarray}
  \rho(0,x^i) & = & Y_{\ell m}\, A \exp \left[ - \frac{(r-R)^2}{W^2}
  \right] \text{,}
\end{eqnarray}
where we choose the multipole $\ell=2,\, m=2$, the amplitude $A=1$,
the radius $R=20$, and the width $W=1$.  We use conventions such that
\begin{eqnarray}
  Y_{22} & = & \sqrt{\frac{15}{32\pi}} \sin^2\theta \left( \cos^2\phi
    - \sin^2\phi \right) \text{.}
\end{eqnarray}
We impose $u=0$, $\rho=0$, and $v_i=0$ with the  penalty method as
outer boundary conditions.  This means that this condition is imposed
onto the incoming characteristic modes.
Since the inner boundary is an outflow
boundary, no boundary condition is imposed there.  At
the outer boundary, $\rho$ is indistinguishably close to zero at $t=0$
(much closer than the floating point round-off error), so that there is
no noticeable discontinuity to the initial condition.
We use again the penalty strength $\delta=0$ for both
inter-patch and outer boundaries.

We use the patch-local tensor basis for this example.  We use $21$
grid points in the angular directions and $1001$ grid points in the
radial direction.  We use here ---for no particular reason---
different discretisation parameters.  It is our experience that the
stability of the system does not depend on the particular choice of
stencil, as long as it satisfies summation by parts \cite{Diener05b}.
We use here the $D_{8-4}$ stencil, which is globally fifth order
accurate, and add compatible MSN artificial dissipation to the system. We
choose a dissipation coefficient $\epsilon=0.2$, and we do not scale the
dissipation with the grid spacing $h$.  The overall system is then
fifth order accurate.  We use a fixed time step size $\Delta t=0.05$
with a fourth order accurate Runge--Kutta integrator.

\begin{figure*}
  \includegraphics[width=\textwidth]{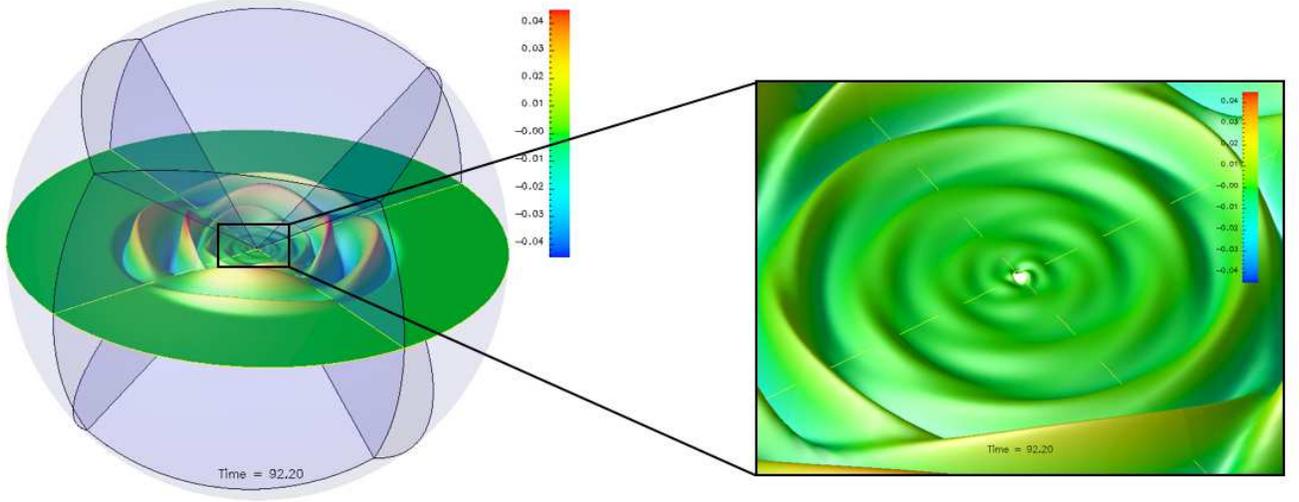}
  \caption{The patch system and the scalar wave configuration at
    $t=92.2$, also enlarging the region near the excision boundary
    inside the horizon.  The background is a rotating black hole with
    $a=0.9$, the initial condition is an $\ell=2,\, m=2$ multipole.
    Note the
    large scale difference between the outer and the inner boundary,
    which is handled ``naturally'' and without mesh refinement.  There
    are no artifacts visible at the inter-patch boundaries.}
  \label{fig:Mover5K-green}
\end{figure*}

\begin{figure*}
  \subfigure[$\ell=2,\, m=2$]
  {\includegraphics[width=0.45\textwidth]{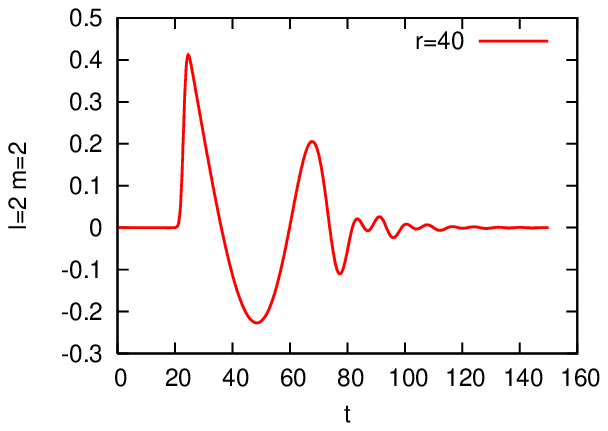}}
  \subfigure[$\ell=4,\, m=2$]
  {\includegraphics[width=0.45\textwidth]{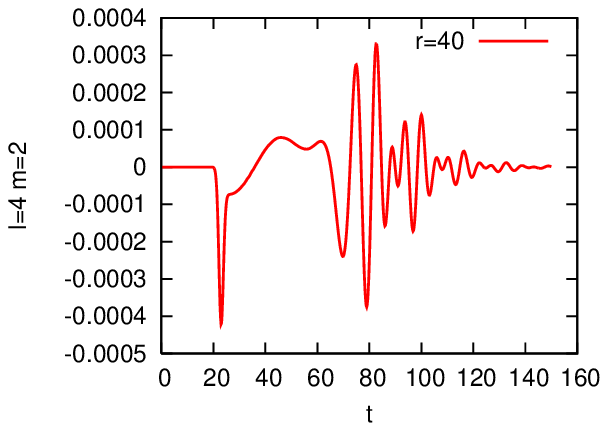}}
  
  \caption{The $\ell=2,\, m=2$ and the $\ell=4,\, m=2$ modes,
    extracted at $r=40$.  The $\ell=4,\, m=2$ mode is excited through
    mode--mode coupling.  Its amplitude is $10^3$ times smaller than
    the $\ell=2,\, m=2$ mode.  All other modes with $\ell \le 4$ are
    zero up to floating point round-off error.}
  \label{fig:Mover5K-r3}
\end{figure*}

Figure \ref{fig:Mover5K-green} shows a snapshot of the simulation at
$t=92.2$.  At that time, the wave pulse has traveled approximately
half the distance to the outer boundary.  The inter-patch boundaries
are smooth, although the configuration is not axisymmetric.  Figure
\ref{fig:Mover5K-r3} shows extracted wave forms from this simulation
for the $\ell=2,\, m=2$ and for the $\ell=4,\, m=2$ modes.  The
$\ell=2$ mode is present in the initial condition.
The $\ell=4$ mode is excited through mode--mode coupling.  Its
amplitude is $10^3$ times smaller than the $\ell=2$ mode at this time.
The $\ell=4$ mode has converged at this resolution, i.e., it does not
change noticeably when the resolution is changed.

We have also simulated initial data consisting of an $\ell=2,\, m=0$
mode.  In this case, both the $\ell=0,\, m=0$ and the $\ell=4,\, m=0$
modes are excited through mode--mode coupling.  As expected, the
mode--mode coupling vanishes when the spin of the background spacetime
is set to zero.

%% \begin{figure}
%%   \includegraphics[width=0.45\textwidth]{runs/scalar_wave_vertical}
%%   \caption{The same picture as \ref{fig:Mover5K-green}, but vertical.}
%% \end{figure}

We determine the complex quasi normal frequency $\omega = \omega_R + i
\omega_I$ from the extracted wave form of the $\ell=2,\, m=2$ mode.
We fit the wave seen by an observer at radius $r=5$ to a function
\begin{eqnarray}
  f(t) & = & A \sin(\omega_R t - \phi) \exp(\omega_I t) \text{.}
\end{eqnarray}
This fit is performed for the real and imaginary part of the complex
frequency as well as for the amplitude $A$ and a phase $\phi$.  Table
\ref{tab:frequencies} shows the frequencies we obtain from our simulations for
the $\ell=2,\, m=2$ and $m=-2$ modes and we compare the predictions
made by perturbation theory \cite{Berti:2004um,
  Berti:2005eb}.  A detailed study of quasi-normal mode frequencies and
excitation coefficients of scalar perturbations of Kerr black holes is
in preparation \cite{Dorband05a}.

For reasons of comparison between a code using the global tensor basis
and one using the patch-local tensor basis, we evolved a similar
physical system using both of these methods.  We now choose a spin of
$a=0.5$ and initial data with an $\ell=2,\, m=0$ angular dependency.
The frequencies obtained by both codes, together with the predictions from
perturbation theory are shown in table \ref{tab:frequencies}.
We find that the choice of tensor
basis has little influence on the accuracy of the results.

\begin{table*}[htdp]
  \begin{tabular}{lll|l|lll|llll}
    \multicolumn{3}{c|}{Mode} & \multicolumn{1}{c|}{Spin} & \multicolumn{3}{c|}{$\omega_{\mathrm{peturbation}}$} & \multicolumn{4}{c}{$\omega_{\mathrm{numerical}}$} \\\hline
    $l=2$ & $,$ & $m=2$  & $0.9$ & $0.781638$ & $-$ & $0.0692893 i$ & $0.796527 (030)$ & $-$ & $0.0680891 (010) i$ \\
    $l=2$ & $,$ & $m=-2$ & $0.9$ & $0.387710$ & $-$ & $0.0935902 i$ & $0.387678 (001)$ & $-$ & $0.0934718 (100) i$ \\
    $l=2$ & $,$ & $m=0$  & $0.5$ & $0.491962$ & $-$ & $0.094630  i$ & $0.491824 (100)$ & $-$ & $0.0946523 (200) i$ & (local) \\
    $l=2$ & $,$ & $m=0$  & $0.5$ & $0.491962$ & $-$ & $0.094630  i$ & $0.492432 (001)$ & $-$ & $0.0944723 (300) i$ & (global)
  \end{tabular}
  \caption{Comparison of the scalar quasi normal frequencies obtained
    with the multi block method using a global and local tensor basis,
    and the values predicted by perturbative methods
    \cite{Berti:2004um, Berti:2005eb}.  The given error estimates for
    the numerical values come from the uncertainty induced by the
    fitting procedure.  Details about that can be found in
    \cite{Dorband05a}.}
  \label{tab:frequencies}
\end{table*}

\section{Evolving the vacuum Einstein equations}
\label{sec:einstein}

We evolve the vacuum Einstein equations using the symmetric
hyperbolic formulation introduced in \cite{Sarbach02b}.
It includes
as variables the three metric $g_{ij}$, the extrinsic curvature
$K_{ij}$, the lapse $\alpha$, and extra variables denoted by $d_{kij}$
and $A_i$.  When all the constraints are satisfied, these are related
to the three-metric and lapse by $d_{kij} = \partial_k g_{ij}$ and
$A_k = \partial_k \ln\alpha$.  The formulation admits any of the
Bona--Masso slicing conditions while still being symmetric hyperbolic.
The shift has to be specified in advance as an arbitrary function of
the spacetime coordinates $t$ and $x^i$.  In the tests below, we use a
time harmonic slicing condition and a time-independent shift.  The
characteristic modes and speeds are listed in \cite{Sarbach02b}.

Previous $3D$ black hole simulations using this formulation were
presented in \cite{Tiglio2003b}, using a low order Cartesian code
and cubic excision.  In \cite{Sarbach04}, constraint-preserving
boundary conditions for this formulation were constructed; this paper
then studies the well-posedness of the resulting initial-boundary
value problem, and tests a numerical implementation of those boundary
conditions in fully non-linear $3D$ scenarios as well, again with a
Cartesian code.

After some initial (and quite lengthy) experiments with a patch-local
tensor basis, we decided to use a global tensor basis instead.%
\footnote{We are grateful to O. Sarbach for suggesting and insisting
  on this.}
We
find that patch-local tensor bases increase the complexity of the
inter-patch boundary conditions very much, because the characteristic
decomposition of the field variables needs to be combined with the
tensor basis transformation at the patch boundaries.
Converting the partial derivatives into the global tensor basis
is trivial in comparison.  In addition to that, analysing and
visualising the output of a simulation is also made much easier when a
global tensor basis is used.

\subsection{Robust stability test}

The robust stability test in numerical relativity consists of evolving
featureless initial conditions to which random noise has been added.
This test was initially suggested in \cite{Szilagyi00a, Szilagyi02b}
and later refined in the so-called Mexico tests
\cite{Alcubierre2003:mexico-I};
see also \cite{applesweb1}.  The first stage of this test has a domain
that is periodic in all directions, i.e., has a $T^3$ topology.  The
most difficult stage of the test has a spherical outer boundary
through which noise is injected.

We implement the $T^3$ topology with a single patch, using penalty
inter-patch boundary conditions to give the system a toroidal
topology.  We also use a six-patch topology and set the incoming modes
to Minkowski on both the inner and outer boundary.
The single patches have outer boundaries at $x^i \in
[-1;+1]$, the six-patch system has the inner boundary at $r=1.9$ and
the outer boundary at $r=11.9$.
We use a Minkowski spacetime as background and add random
noise with an amplitude of $10^{-8}$  to all variables.
Since the Minkowski spacetime has no
intrinsic scale, this amplitude should be compared to our floating
point accuracy of approximately $10^{-16}$.  Terms that are
quadratic in the noise amplitude have then the same order of magnitude
as floating point inaccuracies.

In the runs shown below, we choose the penalty parameter $\delta=0.5$.
We discretise the domain with $21^3$, $41^3$, and $81^3$ grid points
per patch.  We use the $D_{8-4}$ derivative operator and its associated
KO dissipation with a parameter $\epsilon = 0.5$. 
We also use an adaptive Runge--Kutta time integrator.  For
comparison, we also show results using a six-patch system with the
$D_{6-5}$ operator, using the same dissipation parameter.

\begin{figure}
  \includegraphics[width=0.45\textwidth]{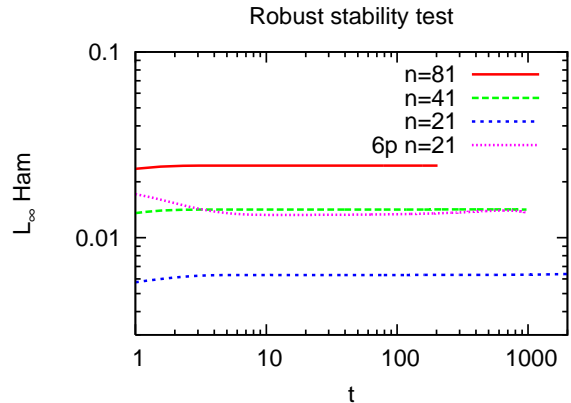}
  \caption{Robust stability test for the Einstein equations.  This
    graph compares three different resolution of a single patch and
    the six-patch system.  Random noise is initially added to all
    variables.  At late times, the Hamiltonian constraint does not
    grow with time; the system is strictly stable.  The two runs
    marked $n=81$ and $6p\; n=21$ have the highest resolutions; we
    aborted them before they reached $t=2000$, which corresponds to
    $1000$ crossing times for the single patch.}
\label{fig:random}
\end{figure}

Figure \ref{fig:random} shows the $L_{\infty}$ norm of the Hamiltonian
constraint as a function of time.  The
constraints remain essentially constant.  Note that the
constraints do not converge to zero with increasing resolution,
since the random data are different for each run and do not have a
continuum limit.

\subsection{One-dimensional gauge wave}
\label{sec:gauge-wave}

One of the most difficult of the Mexico tests
\cite{Alcubierre2003:mexico-I, applesweb1} is the gauge wave test.
This is a one-dimensional
non-linear gauge wave, i.e., flat space in a non-trivial coordinate
system.  This setup lives in a $T^1$ domain, i.e., it has again
periodic boundaries, which we implement either as manifestly
periodic boundary conditions or as penalty inter-patch boundary
conditions.  We use the slightly modified gauge wave which has the
line element
\begin{eqnarray}
  ds^2 & = & -H\,dt^2 + H\,dx^2 + dy^2 + dz^2
\end{eqnarray}
with
\begin{eqnarray}
  H & = & \exp \left[ A \sin \left( \frac{2\pi(x-t)}{L} \right)
  \right] \text{.}
\end{eqnarray}
We choose the wave length $L$ to be the size of our domain, and set
the amplitude $A$ to $0.5$.

\begin{figure}
  \subfigure[Manifestly periodic]
  {\includegraphics[width=0.45\textwidth]{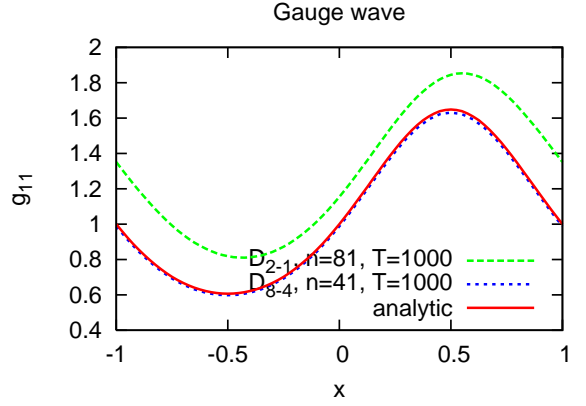}}
  
  \subfigure[Periodic via penalties]
  {\includegraphics[width=0.45\textwidth]{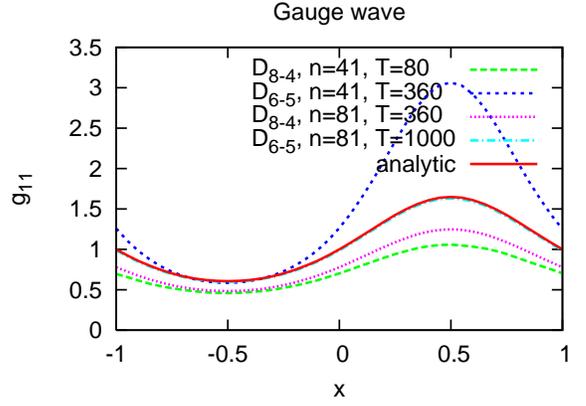}}
  
  \caption{Results for the gauge wave test case, comparing
    differencing operators and resolutions.  With manifestly periodic
    boundaries, the system is both stable and highly accurate after
    $1000$ crossing times.  With periodicity imposed via penalty
    boundary conditions, the system is less accurate, and lower
    resolutions are finally unstable after $g_{11}$ becomes negative.
    However, with sufficient resolution and high order derivatives, our
    system is still highly accurate after $1000$ crossing times.}
  \label{fig:gauge-wave}
\end{figure}

Our domain is one-dimensional with $x \in [-1;+1]$.  Different from
\cite{Alcubierre2003:mexico-I}, we place grid points onto the
boundaries.  We use either $41$ or $81$ grid points.  Figure
\ref{fig:gauge-wave} compares the shape of the wave form at late times
to its initial shape.  Our system is stable and very accurate even
after $1000$ crossing times when we use manifestly periodic boundary
conditions.  When we impose periodicity via penalties, the system is
less accurate, and these inaccuracies lead to a drift which finally
leads to a negative $g_{11}$, which is unstable.  With $81$ grid
points per wave length, however, our system is both stable and very
accurate after $1000$ crossing times even with penalty boundaries.

Standard lore says that a second order discretisation requires one to
use at least 20 grid points per wave length.  This would seem to make
it excessive to use 81 grid points per wave length with our fifth
order scheme.  However, this is not so.  According to Kreiss and
Oliger \cite{Kreiss73}, using 20 grid points per wave length
introduces a 10\% error for each crossing time, making it impossible
to evolve meaningfully for 1000 crossing times.  Achieving a 1\% error
after 1000 crossing time requires about 2000 grid points for a second
order scheme, and approximately 73 grid points for a fourth order
scheme.  Given these numbers, using 81 grid points for our fifth order
scheme is appropriate.

\subsection{Weak gravitational waves}

We now consider perturbations of a flat spacetime, using the
Regge--Wheeler (RW) perturbation theory \cite{Regge57}
to construct an exact solution
to the linearised constraints of Einstein's equations, which we evolve with
the fully non-linear equations.  We linearise
about the Minkowski spacetime, i.e., we choose a background spacetime
with the ADM mass $M=0$.

For simplicity we consider an $\ell=2,\, m=0$ odd parity perturbation.
The resulting metric in the Regge--Wheeler gauge and in spherical
coordinates is
\begin{eqnarray}
  \label{eqn:rw3}
  ds^2 & = & -dt^2 + dr^2 +r^2 (d\theta^2 + \sin^2\theta\, d\phi^2)
  \\\nonumber
  & & {} - 6\, \delta\, r \dot\Psi\, \sin^2\theta\, \cos\theta\, dr\, d\phi
  \\\nonumber
  & & {} - 6\, \delta (\Psi + r\Psi') \sin^2\theta\, \cos\theta\, dt\, d\phi
\end{eqnarray}
where a prime denotes derivative with respect to $r$, and where
$\delta $ is a parameter that determines the ``strength'' of the
perturbation (not to be confused with the penalty term, for which we use the
same symbol).  This metric satisfies the linearised constraints for
{\em any} functions $\Psi(t=0,r)$ and $\dot\Psi(t=0,r)$, and satisfies
all the linearised Einstein equations if $\Psi$ satisfies the
Regge-Wheeler equation
\begin{eqnarray}
  \ddot\Psi & = & \Psi'' - \frac{6}{r^2}\Psi \text{.}
\end{eqnarray}
It is simple to construct a purely outgoing, exact solution to the
previous equation:
\begin{eqnarray}
  \Psi(u,r) & = & \frac{d^2F}{du^2} - \frac{3}{r}\frac{dF}{du} + \frac{3}{r^2}F
\end{eqnarray}
where $F(u)$ is an arbitrary function of $u=r-t$.  The above metric
is very similar to the one in \cite{Teukolsky82}, except
that ours is in the Regge--Wheeler gauge.

However, we use a different coordinate condition for our evolutions.
We construct from the previous metric initial conditions in Cartesian
coordinates for $g_{ij}$, $K_{ij}$, and $d_{kij}$.  We set the initial
lapse to one and evolve it through the time harmonic slicing
condition, and set the shift to zero at all times.

\begin{figure}
  \includegraphics[width=0.45\textwidth]{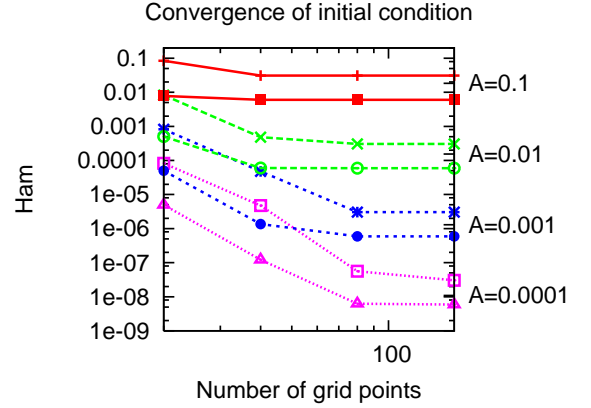}
  \caption{Norms of the Hamiltonian constraint for initial conditions
    with different amplitudes as a function of the number of grid
    points on each patch in each direction.  For each initial
    amplitude we show both the $L_2$ and the $L_{\infty}$ norm.  Since
    the initial conditions only satisfy the linearised constraints, we
    clearly see the non-linear constraint violation at the highest
    resolutions.}
  \label{fig:lin_constraints}
\end{figure}

We now want to check at what point non-linear effects begin to have an
effect on the constraints.  That is, we want to find out what
resolutions are required to see that the full, non-linear constraints
do not actually converge to zero.  As initial condition for the
function $\Psi$ and its time derivative we choose
\begin{eqnarray}
  \Psi(t=0,r) & = & A \exp \left( - \frac{(r-r_0)^2}{\sigma^2} \right)
  \\
  \dot\Psi(t=0,r) & = & -2\, \frac{(r-r_0)}{\sigma^2}\,
   B \exp \left( - \frac{(r-r_0)^2}{\sigma^2} \right) \text{.}
\end{eqnarray}
The non-linear
constraint violations should have an
approximate quadratic dependence on the amplitude parameter (called $A$
below).  Figure \ref{fig:lin_constraints} quantifies this
violation.  It displays the $L_2$ and $L_{\infty}$ norms of the
non-linear Hamiltonian constraint, measured in local coordinates, for
different families of initial conditions as a function of resolution.  We
use the seven-patch system described in section
\ref{sec:patch-systems} with the outer boundary at $r=6$, while the
inner, cubic patch has an extent of $\pm 1$.  The initial condition
parameters are $B=-A$, $\sigma = 0.6$, and $r_0=3$, with amplitudes
$\delta=10^{-1}, 10^{-2}, 10^{-3}, 10^{-4}$.  We use the $D_{6-5}$
derivative with $N^3$ grid points on each patch for $N=21,41,81,161$.
At the highest resolution, the numerical constraints reach their
non-zero continuum values.  This highest resolution corresponds to
$24$ grid points per $\sigma$ in the radial direction.  This number is
comparable to the \emph{coarsest} resolutions that we use in the
simulations presented below.  This means that those simulations use
constraint-violating initial conditions, and cannot expect the
constraints to decrease with increasing resolution.

\begin{figure}
  \includegraphics[width=0.45\textwidth]{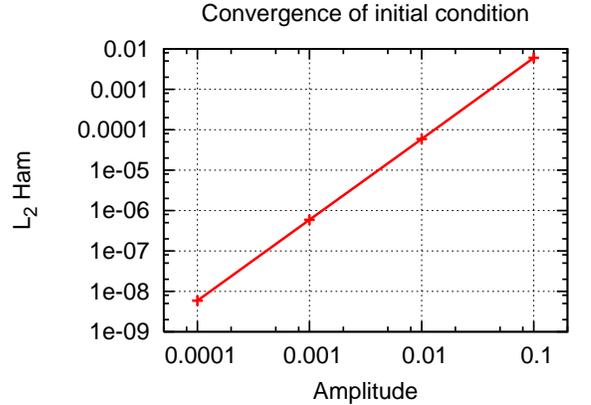}
  \caption{Hamiltonian constraint violation in the $L_2$ norm as a
    function of the amplitude of the perturbation for the highest
    resolution.  This shows the expected quadratic dependence on the
    amplitude.}
  \label{fig:quad}
\end{figure}

\begin{figure}
  \includegraphics[width=0.45\textwidth]{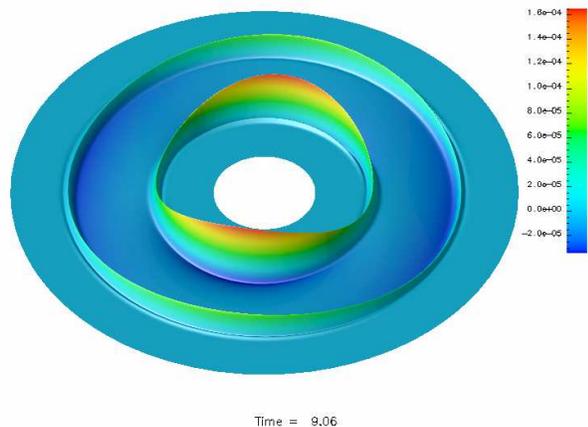}
  \caption{Evolution of weak gravitational waves.  This shows the
    component $K_{xx}$ of the extrinsic curvature in the equatorial
    plane at $t=9.06$.  The gravitational wave packet started as a
    spherical shell approximately in the middle between the inner and
    outer boundaries, and has then split into two packets which travel
    outwards and inwards, respectively.}
  \label{fig:gr_waves}
\end{figure}

The non-linear constraint violation should be approximately a
quadratic function of the amplitude of the perturbation $\delta$, at
least for small values of $\delta$.  Figure \ref{fig:quad} shows that
this is indeed the case.  It displays the Hamiltonian constraint
violation in the $L_2$ norm for the highest resolution of the previous
figure as a function of the amplitude $\delta$.  The measured slope is
$2.0002$.  Figure \ref{fig:gr_waves} shows a sample
evolution of this odd parity initial condition family, using the
six-patch geometry.

In order to evaluate the accuracy of our code we now choose an initial
condition corresponding to an exact solution of the outgoing type
described above.  We do so by choosing
\begin{eqnarray}
  F(u) & = & \exp \left( - \frac{(r-r_0)^2}{\sigma^2} \right) \text{.}
\end{eqnarray}
In these evolutions the parameters that determine the initial condition are
$\sigma=1$, $r_0=30$, and amplitude $\delta=10^{-3}$, with inner
and outer boundaries at $r=10$ and $r=60$, respectively.

\begin{figure}
  \subfigure
  {\includegraphics[width=0.45\textwidth]{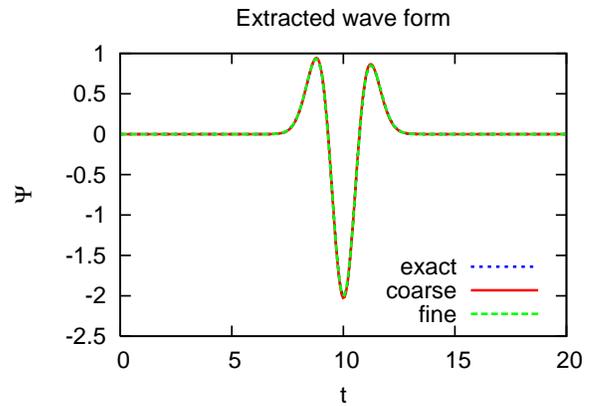}}
  \subfigure
  {\includegraphics[width=0.45\textwidth]{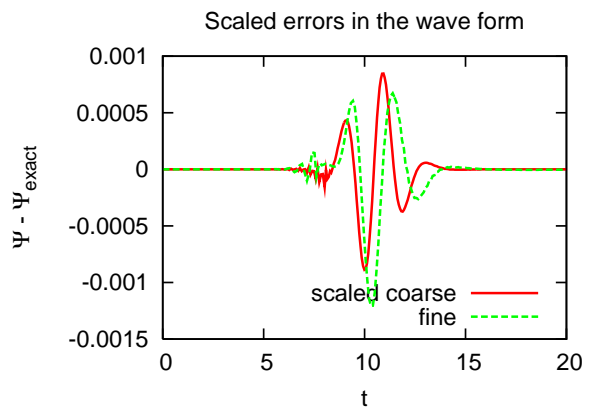}}
  \caption{The upper panel shows the Regge--Wheeler function $\Psi$
    vs.\ time, comparing the numerical and the exact solution at
    $r=40$.  The agreement is excellent, especially for a
    three-dimensional, fully nonlinear code.  The lower panel shows
    the scaled differences to the exact solution for two resolutions
    with $16\times 16\times 1001$ and $22\times 22\times 1401$ grid
    points.  This
    demonstrates fifth order convergence, as should be the case for the
    $D_{8-4}$ operator.}
  \label{fig:gr_waves2}
\end{figure}

We extract the wave forms from the numerical results by calculating
the Regge--Wheeler function at each grid point.  We then average over
one radial shell.  There is no need for interpolating to a sphere,
which simplifies the extraction procedure greatly, and probably also
improves its accuracy.
Figure \ref{fig:gr_waves2} shows the numerically extracted $\Psi_e$ at
$r=40$, and compared to its exact value $\Psi$, for an evolution using
the $D_{8-4}$ derivative, with the dissipation parameter
$\epsilon=0.05$.  We used two resolutions with $16\times 16\times
1001$ and $22\times 22\times 1401$ grid points on each patch.
The agreement is excellent.

\section{Conclusions}
\label{sec:conclusions}

We have motivated the use of multi-patch systems in general
relativity, and have described a generic infrastructure for
multi-patch time evolutions.  Their main advantages are smooth
boundaries and constant angular resolution, which makes them very
efficient for representing systems requiring high resolution in the
centre and having a radiative zone far away.  They may even render
fixed mesh refinement unnecessary in many cases.

We use the penalty method for inter-patch boundary conditions.  It
would equally be possible to use e.g.\ interpolation between the
patches.  A direct comparison of these different approaches would be
very interesting.  Our evolution systems are first order symmetric
hyperbolic, but second order systems could be used as well.  We use
this infrastructure with high-order finite differencing operators, but
other discretisations such as e.g.\ pseudo-spectral collocation
methods can also be used.  Our infrastructure is based on Cactus and
Carpet and runs efficiently in parallel.

We have discussed the relative advantages of using global and
patch-local tensor bases, and we have compared the accuracy and
stability of both approaches.  We suggest that using a global basis is
substantially more convenient, both in the implementation of the code
and in the post-processing of the generated output.

We have tested this infrastructure with a scalar wave equation on a
fixed, stationary background, and with a symmetric hyperbolic
formulation of the Einstein equations.  We have shown that our
multi-patch system with penalty boundary conditions is robustly stable
and can also very accurately reproduce the nonlinear gauge wave of the
\emph{Apples with Apples} tests, which has been the most difficult of
these tests for other codes.

Finally, we have simulated three-dimensional weak gravitational waves
in three dimensions, using the same nonlinear code, and have
accurately extracted the gravitational radiation.  The latter is made
especially simple since the wave extraction spheres are aligned with
the numerical grid.

We believe that multi-patch systems, which provide smooth boundaries,
will be an essential ingredient for discretising well-posed initial
boundary value problems.

\begin{acknowledgments}
  We thank Olivier Sarbach for suggesting to use a global tensor
  basis; our work would have been much more complicated if we had not
  taken this route.  We thank Jonathan Thornburg for many inspiring
  discussions and helpful hints about multiple grid patches.  We also
  engaged in valuable discussions with Burkhard Zink, especially
  regarding penalty inter-patch conditions for nonlinear equations,
  and with Emanuele Berti regarding scalar perturbations of a Kerr
  spacetime.
  We thank Enrique Pazos for his contributions to the design of our
  Regge--Wheeler code.
  We are very grateful to Frank Muldoon for his suggestions about
  automatic grid generation, and for the grids he generated for us.
  We thank Ian Hawke, Christian Ott, Enrique Pazos, Olivier Sarbach,
  and Ed Seidel for proofreading a draft of this paper.
  
  During our work on this project, we enjoyed hospitality at Cornell
  University, at the Cayuga Institute of Physics, at the
  Albert--Einstein--Institut, and at the Center for Computation \&
  Technology at LSU.
  
  As always, our numerical calculations would have been impossible
  without the large number of people who made their work available to
  the public: we used the Cactus Computational Toolkit
  \cite{Goodale02a, cactusweb1} with a number of locally developed
  thorns, the LAPACK \cite{laug, lapackweb} and BLAS \cite{blasweb}
  libraries from the Netlib Repository \cite{netlibweb}, and the LAM
  \cite{burns94:_lam, squyres03:_compon_archit_lam_mpi, lamweb} and
  MPICH \cite{Gropp:1996:HPI, mpich-user, mpichweb} MPI \cite{mpiweb}
  implementations.
  
  Computations for this work were performed on Peyote at the AEI, on
  Helix, Santaka, and Supermike at LSU, at NCSA under allocation
  MCA02N014, and at NERSC.
  
  This work was partially supported by the SFB/TR-7 ``Gravitational
  Wave Astronomy'' of the DFG, the NSF under grant PHY0505761 and NASA
  under grant NASA-NAG5-1430 to Louisiana State University, and by the
  NSF under grants PHY0354631 and PHY0312072 to Cornell University.
  This work was also supported by the Albert--Einstein--Institut,
  Cornell University, by the Horace Hearne Jr.\ Institute for
  Theoretical Physics at LSU, and by the Center for Computation \&
  Technology at LSU.
\end{acknowledgments}

\bibliographystyle{bibtex/apsrev-titles}
\bibliography{bibtex/references}

\end{document}